\documentclass[12pt]{article}

\usepackage{graphicx}
\usepackage{amsmath,amsfonts,amssymb}
\usepackage{hyperref}

\newcommand{\bC}{{\bf C}}

\title{  {\tt ClustGeo}: an R package for hierarchical clustering with spatial constraints}

\author{
 Marie Chavent
	 \thanks{Universit\'e de Bordeaux,}
	 \thanks{Inria Bordeaux Sud-Ouest,}
	 \thanks{Institut de Math\'ematiques de Bordeaux,}
	 \hspace{0,2em}
  \and Vanessa Kuentz-Simonet
   	 \thanks{IRSTEA, UR ETBX, Centre de Bordeaux,}
  	 \hspace{0,2em}   
  \and Amaury Labenne 
   	 \footnotemark[4]
  	\hspace{0,2em} 
  \and J\'er\^ome Saracco
  	 \thanks{ENSC - Bordeaux INP.}
	 \footnotemark[2]
	 \footnotemark[3]
}

\date{\today}

\begin{document}

\maketitle

%%%%%%%%%%%%%%%%%%%%%%%%%%%%%%%%%%%%%%%%%%%%

\begin{abstract}
In this paper, we propose a Ward-like hierarchical clustering algorithm including spatial/geographical constraints. Two dissimilarity matrices $D_0$ and $D_1$ are inputted, along with  a mixing parameter $\alpha \in [0,1]$. The dissimilarities can be non-Euclidean and the weights of the observations can be non-uniform. The first matrix  gives the dissimilarities in the  ``feature space'' and the second matrix gives the dissimilarities in the ``constraint space''. The criterion minimized at each stage  is a convex combination of the homogeneity criterion calculated with $D_0$ and the homogeneity criterion calculated with $D_1$. The idea is then  to determine a value of $\alpha$ which increases the spatial contiguity without deteriorating too much the quality of the solution based on the variables of interest i.e. those of the feature space. This procedure is illustrated on a real dataset using the R package \texttt{ClustGeo}.  
\end{abstract}


%%%%%%%%%%%%%%%%%%%%%%%%%%%%%%%%%%%%%%%%%%%%

%%%%%%%%%%%%%%%%%%%%%%%%%%%%%%%%%%%%%%%%%%%%
%%%%%%%%%%%%%%%%%%%%%%%%%%%%%%%%%%%%%%%%%%%%
%%%%%%%%%%%%%%%%%%%%%%%%%%%%%%%%%%%%%%%%%%%%
\section{Introduction}
\label{intro}
%%%%%%%%%%%%%%%%%%%%%%%%%%%%%%%%%%%%%%%%%%%%
%%%%%%%%%%%%%%%%%%%%%%%%%%%%%%%%%%%%%%%%%%%%
%%%%%%%%%%%%%%%%%%%%%%%%%%%%%%%%%%%%%%%%%%%%

The difficulty of clustering a set of $n$ objects into $k$ disjoint clusters is one that is well known among researchers.  Many methods have been proposed either to find the best partition according to a dissimilarity-based homogeneity criterion, or to fit a mixture model of multivariate distribution function.  However,  in some clustering problems, it is relevant to impose constraints on the set of allowable solutions. In the literature, a variety of different solutions have been suggested and applied in a number of fields, including earth science, image processing, social science, and - more recently -  genetics.  The most common type of constraints are contiguity constraints (in space or in time). Such restrictions occur when  the objects in a cluster are required not only to be similar to one other, but also to comprise a  contiguous set of objects.  But what is a contiguous set of objects? 

Consider first that the contiguity between each pair of objects is given by a matrix $\bC=(c_{ij})_{n\times n}$, where $c_{ij}=1$ if the $i$th and the $j$th objects are regarded as contiguous, and 0 if they are not. A cluster $C$  is then considered to be contiguous if there is a path between every pair of objects in $C$ (the subgraph is connected).  Several classical clustering algorithms have been modified to take  this type of constraint into account (see e.g., Murtagh 1985a; Legendre and Legendre 2012;  B\'ecue-Bertaut et al. 2014). Surveys  of some of these methods can be found in Gordon (1996) and Murtagh (1985b). For instance, the  standard hierarchical procedure based on Lance and Williams formula (1967) can be constrained by merging only contiguous clusters at each stage. But what defines ``contiguous" clusters? Usually,  two clusters are regarded as contiguous if there are two objects, one from each cluster, which are linked in the contiguity matrix. But this can lead to reversals (i.e. inversions, upward branching in the tree) in the hierarchical classification. It was proven  that only the complete link algorithm is guaranteed to produce no reversals when relational constraints are introduced in the ordinary hierarchical clustering procedure (Ferligoj  and Batagelj 1982).  Recent implementation of strict constrained clustering procedures are available in the R package  \texttt{const.clust} (Legendre 2014) and in the Python library \texttt{clusterpy} (Duque et al. 2011). Hierarchical clustering of SNPs (Single Nucleotide Polymorphism) with strict adjacency constraint is also proposed in Dehman et al. (2015) and implemented in the R package \texttt{BALD} (www.math-evry.cnrs.fr/logiciels/bald). The recent R package  \texttt{Xplortext} (B\'ecue-Bertaut et al 2017) implements also chronogically constrained agglomerative hierarchical clustering for the analysis of textual data. 

The previous procedures which impose  strict contiguity may separate objects which are very similar into different clusters, if they are spatially apart.  Other non-strict constrained procedures have then been developed, including those referred to as soft contiguity or spatial constraints.   
For example, Oliver and Webster (1989) and Bourgault  et al. (1992) suggest running clustering algorithms on a modified dissimilarity matrix.  This dissimilarity matrix is a combination of the matrix of geographical distances and the dissimilarity matrix computed from non-geographical variables.  According to the weights given to the geographical dissimilarities in this combination, the solution will have more or less spatially contiguous clusters. However,  this approach raises the problem of defining weight in an objective manner.   

In image processing, there are many approaches for image segmentation including for instance usage of convolution and wavelet  transforms.  In this field non-strict spatially constrained clustering methods have been also developed. Objects are pixels and the most common choices for the neighborhood graph are the four and eight neighbors graphs.  A contiguity matrix $\bC$ is used (and not a geographical dissimilarity matrix as previously) but the clusters are not strictly contiguous, as a cluster of pixels does not necessarily represent a single region on the image. Ambroise et al. (1997, 1998)  suggest a clustering algorithm for Markov random fields based on an EM (Expectation-Maximization) algorithm. This algorithm maximizes a penalized likelihood criterion and the regularization parameter gives more or less weight to the spatial homogeneity term (the penalty term).  Recent implementations of  spatially-located data clustering algorithms are available  in  SpaCEM3 (spacem3.gforge.inria.fr), dedicated to Spatial Clustering with EM and Markov Models. This software uses the model proposed in Vignes  and Forbes (2009) for gene clustering via integrated Markov models. In a similar vein, Miele et al. (2014) proposed a model-based spatially constrained method for the clustering of ecological networks. This method embeds geographical information within an EM regularization framework by adding some constraints to the maximum likelihood estimation of parameters. The associated R package is available at http://lbbe.univ-lyon1.fr/geoclust. Note that all these methods are partitioning methods and that the constraints are neighborhood constraints.    

In this paper, we propose a  hierarchical  clustering (and not partitioning) method including  spatial constraints (not necessarily neighborhood constraints). This Ward-like algorithm uses two dissimilarity matrices $D_0$ and $D_1$ and a mixing parameter $\alpha \in [0,1]$. The dissimilarities are not necessarily Euclidean (or non Euclidean) distances and the weights of the observations can be non-uniform. The first matrix  gives the dissimilarities in the  `feature space' (socio-economic variables or grey levels for instance). The second matrix gives the dissimilarities in the `constraint space'. For instance, $D_1$ can be a matrix of geographical distances or a  matrix built from the contiguity matrix $\bC$.  The mixing parameter $\alpha$ sets the importance of the constraint in the clustering procedure. The criterion minimized at each stage  is a convex combination of the homogeneity criterion calculated with $D_0$ and the homogeneity criterion calculated with $D_1$. The parameter $\alpha$ (the weight of this convex combination) controls the weight of the constraint in the quality of the solutions. When $\alpha$ increases, the homogeneity calculated with $D_0$ decreases whereas the homogeneity calculated with $D_1$ increases. The idea is to determine a value of $\alpha$ which increases the spatial-contiguity without deteriorating too much the quality of the solution on the variables of interest. The R package \texttt{ClustGeo}  (Chavent et al. 2017)  implements this constrained hierarchical clustering algorithm and a procedure for the choice of $\alpha$.

The paper is organized as follows. After a short introduction (this section), Section \ref{sec:2} presents the criterion optimized when the Lance-Williams (1967) parameters are used in Ward's minimum variance method  but dissimilarities  are not necessarily Euclidean (or non-Euclidean) distances.
We  also show how to implement this procedure with the package {\tt ClustGeo} (or the R function  {\tt hclust}) when non-uniform weights are used. In Section \ref{sec:3} we present the constrained hierarchical clustering algorithm which optimizes a convex combination of this criterion calculated with two dissimilarity matrices. Then the procedure for the choice of the mixing parameter is presented as well as a description of the functions implemented in the package  {\tt ClustGeo}. In Section \ref{sec:4} we illustrate the proposed hierarchical clustering process with geographical constraints using the package  {\tt ClustGeo}  before a brief discussion given in Section \ref{sec:5}. 

Throughout the paper, a real dataset is used for illustration and reproducibility purposes. This dataset contains 303 French municipalities described based on four socio-economic variables. The matrix $D_0$ will contain the socio-economic distances between municipalities and the matrix $D_1$ will contain the geographical distances. The results will be easy to visualize on a map.

%%%%%%%%%%%%%%%%%%%%%%%%%%%%%%%%%%%%%%%%%%%%
%%%%%%%%%%%%%%%%%%%%%%%%%%%%%%%%%%%%%%%%%%%%
%%%%%%%%%%%%%%%%%%%%%%%%%%%%%%%%%%%%%%%%%%%%
\section{Ward-like  hierarchical clustering with dissimilarities and non-uniform weights}
\label{sec:2}
%%%%%%%%%%%%%%%%%%%%%%%%%%%%%%%%%%%%%%%%%%%%
%%%%%%%%%%%%%%%%%%%%%%%%%%%%%%%%%%%%%%%%%%%%
%%%%%%%%%%%%%%%%%%%%%%%%%%%%%%%%%%%%%%%%%%%%

Let us consider a set of $n$ observations. Let $w_i$ be the weight of the $i$th observation for $i=1,\dots,n$.
Let $D=[d_{ij}]$ be a $n\times n$ dissimilarity matrix associated with the $n$ observations, where $d_{ij}$ is the dissimilarity measure between observations $i$ and $j$. 
Let us recall that the considered dissimilarity matrix $D$ is not necessarily a matrix of Euclidean (or non-Euclidean) distances. When $D$ is not a matrix of Euclidean distances, the usual inertia criterion (also referred to as variance criterion) used in Ward (1963) hierarchical clustering approach   is meaningless and the Ward algorithm implemented with the Lance and Williams (1967) formula has to be re-interpreted.  
The Ward method  has already been generalized to use with non-Euclidean distances, see e.g. Strauss and von Maltitz (2017) for $l_1$ norm or Manhattan distances.   In this section the more general case of dissimilarities is studied.
We first present the homogeneity criterion which is optimized in  that case and the underlying  aggregation measure which leads to a Ward-like hierarchical clustering process. We then provide an illustration using the package  {\tt ClustGeo} and the well-known R function  {\tt hclust}.

\subsection{The Ward-like method}

\paragraph{Pseudo-inertia.}
Let us consider a partition $\mathcal{P}_K=(\mathcal{C}_1,\ldots,\mathcal{C}_K)$ in $K$ clusters. The  pseudo-inertia of a cluster  $\mathcal{C}_k$ generalizes the inertia to the case of dissimilarity data (Euclidean or not) in the following way :
\begin{equation}\label{inertia}
I(\mathcal{C}_k)=\sum_{i \in \mathcal{C}_k}\sum_{j \in \mathcal{C}_k}\frac{w_i w_j}{2\mu_k}d^2_{ij} 
\end{equation}
where $\mu_k=\sum_{i \in \mathcal{C}_k}w_i$ is the weight of $\mathcal{C}_k$. The smaller the pseudo-inertia $I(\mathcal{C}_k)$  is, the more homogenous are the observations belonging to the cluster $\mathcal{C}_k$.

\noindent
 The  pseudo within-cluster inertia of the partition $\mathcal{P}_K$ is therefore:
$$W(\mathcal{P}_K)=\sum_{k=1}^K I(\mathcal{C}_k).$$
The smaller this pseudo within-inertia $W(\mathcal{P}_K)$ is, the more homogenous is the partition in $K$ clusters.

\smallskip

\paragraph{Spirit of the Ward hierarchical clustering.}
To obtain a new partition  $\mathcal{P}_{K}$ in $K$ clusters from a given partition $\mathcal{P}_{K+1}$ in $K+1$ clusters,   
the idea is to aggregate the two clusters $\mathcal{A}$ and $\mathcal{B}$ of  $\mathcal{P}_{K+1}$ such that the new partition has minimum within-cluster inertia (heterogeneity, variance), that is:
\begin{equation}\label{opti1}
\arg\min_{\mathcal{A},\mathcal{B}\in \mathcal{P}_{K+1}} W(\mathcal{P}_{K}), 
\end{equation}
where $\mathcal{P}_{K}=\mathcal{P}_{K+1} \backslash \{\mathcal{A},\mathcal{B}\} \cup \{\mathcal{A}\cup\mathcal{B}\}$ and
$$W(\mathcal{P}_{K})=W(\mathcal{P}_{K+1})-I(\mathcal{A})-I(\mathcal{B})+I(\mathcal{A} \cup \mathcal{B}).$$
Since $W(\mathcal{P}_{K+1})$ is fixed for a given partition $\mathcal{P}_{K+1}$, the optimization problem (\ref{opti1}) is equivalent to: 
\begin{equation}\label{opti2}
\min_{\mathcal{A},\mathcal{B}\in \mathcal{P}_{K+1}} I(\mathcal{A} \cup \mathcal{B})-I(\mathcal{A})-I(\mathcal{B}).
\end{equation}
The optimization problem  is therefore achieved by defining 
$$\delta(\mathcal{A},\mathcal{B}):=I(\mathcal{A} \cup \mathcal{B})-I(\mathcal{A})-I(\mathcal{B})$$
as the aggregation measure between two clusters which is minimized at each step of the hierarchical clustering algorithm.
Note that $\delta(\mathcal{A},\mathcal{B})=W(\mathcal{P}_K)-W(\mathcal{P}_{K+1})$ can be seen as the increase of  within-cluster inertia (loss of homogeneity).

\medskip

\paragraph{Ward-like hierarchical clustering process for non-Euclidean dissimilarities.}
The interpretation of the Ward  hierarchical clustering process  in the case of dissimilarity data is the following: 
\begin{itemize}
\item Step $K=n$: initialization.

The initial partition $\mathcal{P}_n$ in $n$ clusters (i.e. each cluster only contains an observation) is unique.

\item Step $K=n-1,\dots,2$: obtaining the partition in $K$ clusters from the partition in $K+1$ clusters.

At each step $K$, the algorithm aggregates  the two clusters  $\mathcal{A}$ and $\mathcal{B}$ of  $\mathcal{P}_{K+1}$ according to the optimization problem (\ref{opti2}) such that the increase of  the pseudo within-cluster inertia is minimum for the selected partition over the other ones in $K$ clusters. 

\item Step $K=1$: stop. The partition $\mathcal{P}_1$ in one cluster (containing the $n$ observations) is obtained.
\end{itemize}

The hierarchically-nested set of such partitions $\{\mathcal{P}_n,\dots,\mathcal{P}_K,\dots,\mathcal{P}_1\}$ is represented graphically by a tree (also called dendrogram) where the height of a cluster $\mathcal{C} = \mathcal{A} \cup \mathcal{B}$  is
$h(\mathcal{C}):=\delta(\mathcal{A},\mathcal{B}).$

\medskip

In practice, the aggregation measures between the new cluster $\mathcal{A} \cup \mathcal{B}$ and any cluster $\mathcal{D}$ of $\mathcal{P}_{K+1}$ are calculated  at each step thanks to the well-known Lance and Williams  (1967) equation:
\begin{equation}\label{LW-equation}
\begin{array}{lcl}
\delta(\mathcal{A} \cup \mathcal{B},\mathcal{D}) & = & \displaystyle \frac{\mu_\mathcal{A} + \mu_\mathcal{D}}{\mu_\mathcal{A} + \mu_\mathcal{B} + \mu_\mathcal{D}}\delta(\mathcal{A},\mathcal{D}) + \frac{\mu_\mathcal{B} + \mu_\mathcal{D}}{\mu_\mathcal{A} + \mu_\mathcal{B} + \mu_\mathcal{D}}\delta(\mathcal{B},\mathcal{D})\\
&&\\
&-&\displaystyle \frac{\mu_\mathcal{D}}{\mu_\mathcal{A} + \mu_\mathcal{B} + \mu_\mathcal{D}}\delta(\mathcal{A},\mathcal{B}).
\end{array}
\end{equation}

In the first step the partition is $\mathcal{P}_n$ and the aggregation measures between the singletons are calculated with
$$\delta_{ij}:=\delta(\{i\},\{j\})=\frac{w_i w_j}{w_i+w_j} d^2_{ij},$$
and stored in the $n \times n$ matrix  $\Delta=[\delta_{ij}]$. For each subsequent step $K$, the  Lance and Williams formula (\ref{LW-equation}) is used to build the corresponding $K\times K$ aggregation matrix.

The hierarchical clustering process described above is thus suited for non-Euclidean dissimilarities and then for non-numerical data.  In this case, it optimises the  pseudo within-cluster inertia criterion (\ref{opti2}). 

\paragraph{Case when the dissimilarities are Euclidean distances.} 
When the dissimilarities are Euclidean distances calculated from a numerical data matrix $X$ of dimension $n \times p$ for instance, the pseudo-inertia of a cluster $\mathcal{C}_k$ defined in (\ref{inertia}) is now equal to the inertia of the observations in $\mathcal{C}_k$:
$$I(\mathcal{C}_k)=\sum_{i \in \mathcal{C}_k} w_i d^2(x_i,g_k)$$
where $x_i\in \Re^p$ is the $i$th row of $X$ associated with the $i$th observation, and $g_k=\frac{1}{\mu_k}\sum_{i \in \mathcal{C}_k} w_i x_i \in R^p$ is the center of gravity of $\mathcal{C}_k$. 
The aggregation measure $\delta(\mathcal{A},\mathcal{B})$ between two clusters is written then as:
$$\delta(\mathcal{A},\mathcal{B})=\frac{\mu_\mathcal{A} \mu_\mathcal{B}}{\mu_\mathcal{A}+\mu_\mathcal{B}}d^2(g_\mathcal{A},g_\mathcal{B}).$$

\subsection{Illustration using the package {\tt ClustGeo}}
Let us examine how to properly implement  this procedure with R. The dataset is made up of $n=303$ French municipalities described based on $p=4$ quantitative variables and is available in the package {\tt ClustGeo}.  A more complete description of the data is provided in Section \ref{sec:data}.

{\small
\begin{verbatim}
> library(ClustGeo)
> data(estuary)
> names(estuary)
[1] "dat"   "D.geo" "map" 
\end{verbatim}
}

\noindent
To carry out Ward hierarchical clustering, the user can use the function {\tt hclustgeo}  implemented in the package  {\tt ClustGeo} taking the dissimilarity matrix $D$ 
(which is an object of class {\tt dist}, i.e. an object obtained with the function {\tt dist} or a dissimilarity matrix  transformed in an object of class {\tt dist} with the function {\tt as.dist})
and the weights $w=(w_1,\dots,w_n)$ of observations as arguments. 

{\small
\begin{verbatim}
> D <- dist(estuary$dat)
> n <- nrow(estuary$dat)
> tree <- hclustgeo(D, wt=rep(1/n,n))
\end{verbatim}
}

\paragraph{Remarks.}
\begin{itemize}
\item The function {\tt hclustgeo}  is a wrapper of the usual function {\tt hclust} with  the following arguments: 
\begin{itemize}
\item {\tt method = "ward.D"},
\item {\tt d =} $\Delta$,
\item {\tt members = w}.
\end{itemize}
For instance, when the observations are all weighted by $1/n$ , the argument {\tt d} must be the matrix $\Delta=\frac{D^2}{2n}$ and not the dissimilarity matrix $D$:

{\small
\begin{verbatim}
> tree <- hclust(D^2/(2*n), method="ward.D")
\end{verbatim}
}

\item As mentioned before, the user can check that the sum of the heights in the dendrogram is equal to the total pseudo-inertia of the dataset:

{\small
\begin{verbatim}
> inertdiss(D, wt=rep(1/n, n)) # the pseudo-inertia of the data 
[1] 1232.769
> sum(tree$height)
[1] 1232.769
\end{verbatim}
}

\item When the weights are not  uniform, the calculation of the matrix $\Delta$ takes a few lines of code and the use of the function {\tt hclustgeo} is clearly  more convenient than {\tt hclust}:
{\small
\begin{verbatim}
> w <- estuary$map@data$POPULATION # non-uniform weights
> tree <- hclustgeo(D, wt=w)
> sum(tree$height)
[1] 1907989
\end{verbatim}
}
versus
{\small
\begin{verbatim}
> Delta <-  D               
> for (i in 1:(n-1)) {
    for (j in (i+1):n) {
      Delta[n*(i-1) - i*(i-1)/2 + j-i] <-
        Delta[n*(i-1) - i*(i-1)/2 + j-i]^2*w[i]*w[j]/(w[i]+w[j])
        }}
> tree <- hclust(Delta, method="ward.D", members=w)
> sum(tree$height)
[1] 1907989
\end{verbatim}
}
\end{itemize}

%%%%%%%%%%%%%%%%%%%%%%%%%%%%%%%%%%%%%%%%%%%%%%%%%%%%%%%%%%
%%%%%%%%%%%%%%%%%%%%%%%%%%%%%%%%%%%%%%%%%%%%%%%%%%%%%%%%%%
%%%%%%%%%%%%%%%%%%%%%%%%%%%%%%%%%%%%%%%%%%%%%%%%%%%%%%%%%%
\section{Ward-like hierarchical clustering with two dissimilarity matrices}
\label{sec:3}
%%%%%%%%%%%%%%%%%%%%%%%%%%%%%%%%%%%%%%%%%%%%%%%%%%%%%%%%%%
%%%%%%%%%%%%%%%%%%%%%%%%%%%%%%%%%%%%%%%%%%%%%%%%%%%%%%%%%%
%%%%%%%%%%%%%%%%%%%%%%%%%%%%%%%%%%%%%%%%%%%%%%%%%%%%%%%%%%

Let us consider again a set of $n$ observations, and let $w_i$ be the weight of the $i$th observation for $i=1,\dots,n$.
Let us now consider that two  $n\times n$ dissimilarity matrices $D_0=[d_{0,ij}]$ and $D_1=[d_{1,ij}]$ are provided. For instance, let us assume that the $n$ observations are municipalities, $D_0$ can be based on a numerical data matrix of $p_0$ quantitative variables measured on the $n$ observations and $D_1$ can be a matrix containing the geographical distances between the $n$ observations.

\smallskip

\noindent
In this section, a Ward-like hierarchical clustering algorithm is proposed. A mixing parameter $\alpha\in [0,1]$ allows the user to set the importance of each dissimilarity matrix in the clustering procedure. More specifically, if $D_1$  gives the dissimilarities in the constraint space, the mixing parameter $\alpha$ sets the importance of the constraint in the clustering procedure and controls the weight of the constraint in the quality of the solutions.

%==================================================================
%==================================================================
\subsection{Hierarchical clustering algorithm with two dissimilarity matrices}
%==================================================================
%==================================================================

For a given value of $\alpha \in [0,1]$, the algorithm works as follows. Note that the partition in $K$ clusters will be hereafter indexed by $\alpha$ as follows:  $\mathcal{P}_{K}^\alpha$.

\paragraph{Definitions.} The {\bf mixed} pseudo  inertia of the cluster $C_k^\alpha$ (called mixed inertia hereafter for sake of simplicity) is defined as
\begin{equation}\label{heterogeneity-cluster}
I_\alpha(\mathcal{C}_k^\alpha) = (1-\alpha) \sum_{i \in \mathcal{C}_k^\alpha}\sum_{j \in \mathcal{C}^\alpha_k}\frac{w_i w_j}{2\mu_k^\alpha}d^2_{0,ij}  + \alpha \sum_{i \in \mathcal{C}_k^\alpha}\sum_{j \in \mathcal{C}_k^\alpha}\frac{w_i w_j}{2\mu_k^\alpha}d^2_{1,ij},
\end{equation}
where $\mu_k^\alpha=\sum_{i \in \mathcal{C}_k^\alpha}w_i$ is the weight of $\mathcal{C}_k^\alpha$ and $d_{0,ij}$ (resp. $d_{1,ij}$) is the normalized dissimilarity between observations $i$ and $j$ in $D_0$ (resp. $D_1$).

\noindent
The {\bf mixed}  pseudo within-cluster inertia (called mixed within-cluster inertia hereafter for sake of simplicity)  of a partition $\mathcal{P}_K^\alpha=(\mathcal{C}_1^\alpha,\dots,\mathcal{C}_K^\alpha)$ is  the sum of the mixed inertia of its clusters:
\begin{equation}\label{heterogeneity-partition}
W_\alpha(\mathcal{P}_K^\alpha)=\sum_{k=1}^K I_\alpha(\mathcal{C}_k^\alpha).
\end{equation}

\smallskip

\paragraph{Spirit of the Ward-like hierarchical clustering.}
As previously, in order to obtain a new partition  $\mathcal{P}_{K}^\alpha$ in $K$ clusters from a given partition $\mathcal{P}_{K+1}^\alpha$ in $K+1$ clusters,   
the idea is to aggregate the two clusters $\mathcal{A}$ and $\mathcal{B}$ of  $\mathcal{P}_{K+1}$ such that the new partition has minimum {\bf mixed} within-cluster inertia.
The optimization problem can now be expressed as follows: 
\begin{equation}\label{opti3}
\arg\min_{ \mathcal{A},\mathcal{B}\in \mathcal{P}_{K+1}^\alpha} I_\alpha(\mathcal{A} \cup \mathcal{B})-I_\alpha(\mathcal{A})-I_\alpha(\mathcal{B}).
\end{equation}

\paragraph{ Ward-like hierarchical clustering process.}
\begin{itemize}
\item Step $K=n$: initialization.

The dissimilarities can be re-scaled between 0 and 1 to obtain the same order of magnitude: for instance,
$$D_0 \leftarrow \frac{D_0}{\max(D_0)} ~~~\mbox{and}~~~ 
D_1 \leftarrow \frac{D_1}{\max(D_1)}.$$
Note that this normalization step can also be done in a different way.

The initial partition $\mathcal{P}_{n}^\alpha=:\mathcal{P}_n$ in $n$ clusters (i.e. each cluster only contains an observation) is unique and thus does not depend on $\alpha$.

\smallskip

\item Step $K=n-1,\dots,2$: obtaining the partition in $K$ clusters from the partition in $K+1$ clusters.

At each step $K$, the algorithm aggregates the two clusters  $\mathcal{A}$ and $\mathcal{B}$ of $\mathcal{P}_{K+1}^\alpha$ 
according to the optimization problem (\ref{opti3})
such that the increase of  the {\bf mixed}  within-cluster inertia is minimum for the selected partition over the other ones in $K$ clusters. 

More precisely, at step $K$, the algorithm aggregates  the two clusters $\mathcal{A}$ and $\mathcal{B}$ such that the corresponding aggregation measure 
$$\delta_\alpha(\mathcal{A},\mathcal{B}):=W_\alpha(\mathcal{P}_{K+1}^\alpha)-W_\alpha(\mathcal{P}_{K}^\alpha)=I_\alpha(\mathcal{A} \cup \mathcal{B})-I_\alpha(\mathcal{A})-I_\alpha(\mathcal{B}) $$
is minimum. 
\smallskip

\item Step $K=1$: stop. The partition $\mathcal{P}_1^\alpha=:\mathcal{P}_1$ in one cluster is obtained.
Note that this partition is unique and thus does not depend on $\alpha$.
\end{itemize}

\noindent
In the dendrogram of the corresponding hierarchy, the value (height) of a cluster $\mathcal{A} \cup \mathcal{B}$  is given by the agglomerative cluster criterion value
$\delta_\alpha(\mathcal{A},\mathcal{B}).$

\medskip

In practice, the Lance and Williams equation (\ref{LW-equation}) remains true in this context (where $\delta$ must be replaced by $\delta_\alpha$).
The aggregation measure between two singletons are written now:
$$\delta_\alpha(\{i\},\{j\})=(1-\alpha)\frac{w_i w_j}{w_i+w_j} d^2_{0,ij} + \alpha\frac{w_i w_j}{w_i+w_j} d^2_{1,ij}. $$
The Lance and Williams equation is then applied to the matrix
$$\Delta_\alpha=(1-\alpha) \Delta_0 + \alpha \Delta_1.$$
where  $\Delta_0$ (resp. $\Delta_1$) is the $n \times n$ matrix of the values $\delta_{0,ij}=\frac{w_i w_j}{w_i+w_j} d^2_{0,ij}$ (resp.  $\delta_{1,ij}=\frac{w_i w_j}{w_i+w_j} d^2_{1,ij}$).

\paragraph{Remarks.} 
\begin{itemize}
\item The proposed procedure is different from applying directly the Ward algorithm to the ``dissimilarity'' matrix obtained via the convex combination $D_\alpha=(1-\alpha) D_0 + \alpha D_1$. The main benefit of the proposed procedure is that the mixing parameter $\alpha$ clearly controls the part of pseudo-inertia due to $D_0$ and $D_1$ in (\ref{heterogeneity-cluster}). This is not the case when applying directly the Ward algorithm to $D_\alpha$ since it is based on a unique pseudo-inertia.

\medskip

\item When $\alpha=0$ (resp. $\alpha=1$), the hierarchical clustering is only based on the dissimilarity matrix $D_0$ (resp. $D_1$). A procedure to determine a suitable value for the mixing parameter $\alpha$ is proposed hereafter, see Section~\ref{criterion-alpha}.   
\end{itemize}

%==================================================================
%==================================================================
\subsection{A procedure to determine a suitable value for the mixing parameter $\alpha$}
\label{criterion-alpha}
%==================================================================
%==================================================================

The key point is the choice of a suitable value for the mixing parameter $\alpha \in [0,1]$.  This parameter  logically depends on the number of clusters $K$ and this logical dependence is an issue when it comes to decide an optimal value for the parameter $\alpha$.  In this paper a practical (but not globally optimal) solution to this issue is proposed: conditioning on $K$ and choosing $\alpha$ that best compromises between loss of socio-economic and loss of geographical homogeneity.  Of course other solutions than conditioning on $K$  could be explored (conditioning on $\alpha$ or defining a global criterion) but these solutions seem to be more difficult to implement in a sensible procedure.

To illustrate the idea of the proposed procedure, let us assume that the dissimilarity matrix $D_1$ contains geographical distances between $n$ municipalities, whereas the dissimilarity matrix $D_0$ contains distances based on a $n\times p_0$ data matrix $X_0$ of $p_0$ socio-economic variables measured on these $n$ municipalities.  An objective of the user could be to determine a value of $\alpha$ which increases the geographical homogeneity of a partition in $K$ clusters  without adversely affecting socio-economic homogeneity.  These homogeneities can be measured using the appropriate pseudo within-cluster inertias.

Let $\beta\in [0,1]$. Let us introduce the notion of proportion of the total {\bf mixed} (pseudo) inertia explained by the partition $\mathcal{P}_{K}^\alpha$ in $K$ clusters:
$$Q_\beta(\mathcal{P}_K^\alpha)=1-\frac{W_\beta(\mathcal{P}_K^\alpha)}{W_\beta(\mathcal{P}_1)} \in [0,1].$$

\paragraph{Some comments on this criterion.}
\begin{itemize} 
\item When $\beta=0$, the denominator $W_0(\mathcal{P}_1)$ is the total (pseudo)  inertia based on the dissimilarity matrix $D_0$ and the numerator is the (pseudo) within-cluster inertia $W_0(\mathcal{P}_{K}^\alpha)$ based on the dissimilarity matrix $D_0$, i.e. only from the socio-economic point of view in our illustration.

The higher the value of the criterion $Q_0(\mathcal{P}_K^\alpha)$, the more homogeneous the partition $\mathcal{P}_{K}^\alpha$  is from the socio-economic point of view (i.e. each cluster $\mathcal{C}_k^\alpha,~k=1,\dots,K$ has a low inertia $I_0(\mathcal{C}_k^\alpha)$ which means that individuals within the cluster are similar).

When the considered partition $\mathcal{P}_K^\alpha$ has been obtained with $\alpha=0$,  the criterion $Q_0(\mathcal{P}_K^\alpha)$ is obviously maximal (since the partition $\mathcal{P}_K^0$ was obtained by using only the dissimilarity matrix $D_0$),  and this criterion will naturally tend to  decrease as $\alpha$ increases from 0 to 1. 

\medskip

\item Similarly, when $\beta=1$, the denominator $W_1(\mathcal{P}_1)$ is the total (pseudo)  inertia based on the dissimilarity matrix $D_1$ and the numerator is the (pseudo) within-cluster inertia $W_1(\mathcal{P}_{K}^\alpha)$ based on the dissimilarity matrix $D_1$, i.e. only from a geographical point of view in our illustration.

Therefore, the higher the value of the criterion $Q_1(\mathcal{P}_K^\alpha)$, the more homogeneous the partition $\mathcal{P}_{K}^\alpha$  from a geographical point of view.

When the considered partition $\mathcal{P}_K^\alpha$ has been obtained with $\alpha=1$,  the criterion $Q_1(\mathcal{P}_K^\alpha)$ is obviously maximal (since the partition $\mathcal{P}_K^1$ was obtained by using only the dissimilarity matrix $D_1$),  and this criterion will naturally tend to  decrease as $\alpha$ decreases from 1 to 0. 

\medskip

\item For a value of $\beta\in]0,1[$, the denominator $W_\beta(\mathcal{P}_1)$ is a total  {\bf mixed} (pseudo)  inertia which can not be easily interpreted in practice, and the numerator $W_\beta(\mathcal{P}_K^\alpha)$ is the {\bf mixed} (pseudo) within-cluster inertia. 
Note that when the considered partition $\mathcal{P}_K^\alpha$ has been obtained with $\alpha=\beta$,  the criterion $Q_\beta(\mathcal{P}_K^\alpha)$ is obviously maximal by construction, and it will tend to decrease as $\alpha$ moves away from $\beta$. 

\medskip

\item Finally, note that this criterion $Q_\beta(\mathcal{P}_K^\alpha)$ is decreasing in $K$. Moreover, $\forall \beta\in [0,1]$, it is easy to see that $Q_\beta(\mathcal{P}_n)=1$
and $Q_\beta(\mathcal{P}_1)=0$. The more clusters there are in a partition, the more homogeneous these clusters are (i.e. with a low inertia).
Therefore this criterion can not be used to select an appropriate number $K$ of clusters.

\end{itemize} 

\paragraph{How to use this criterion to select the mixing parameter $\alpha$.}
Let us focus on the above mentioned case where the user is interested in determining a value of $\alpha$ which increases the geographical homogeneity of a partition in $K$ clusters  without deteriorating too much the socio-economic homogeneity. 
For a given number $K$ of clusters (the choice of $K$ is discussed later), the  idea is  the following:

\begin{itemize}
\item Let us consider a given grid of $J$ values for $\alpha \in [0,1]$: $$\mathcal{G}=\{\alpha_1=0,\alpha_1,\dots,\alpha_J=1\}.$$

For each value $\alpha_j\in\mathcal{G}$, the corresponding partition $\mathcal{P}_K^{\alpha_j}$ in $K$ clusters is obtained using the proposed Ward-like hierarchical clustering algorithm.

\medskip

\item For the $J$ partitions  $\{\mathcal{P}_K^{\alpha_j},~j=1,\dots,J\}$, the criterion $Q_0(\mathcal{P}_K^{\alpha_j})$ is evaluated. The plot of the points $\{(\alpha_j, Q_0(\mathcal{P}_K^{\alpha_j})),~j=1,\dots,J\}$ provides a visual way to observe the loss of socio-economic  homogeneity of the partition $\mathcal{P}_K^{\alpha_j}$  (from the ``pure'' socio-economic partition $\mathcal{P}_K^{0}$) as  $\alpha_j$ increases from 0 to 1.
 
 \medskip
 
 \item Similarly, for the $J$ partitions  $\{\mathcal{P}_K^{\alpha_j},~j=1,\dots,J\}$, the criterion $Q_1(\mathcal{P}_K^{\alpha_j})$ is evaluated. The plot of the points $\{(\alpha_j, Q_1(\mathcal{P}_K^{\alpha_j})),~j=1,\dots,J\}$ provides a visual way to observe the loss of  geographical homogeneity of the partition $\mathcal{P}_K^{\alpha_j}$ (from the ``pure''  geographical partition $\mathcal{P}_K^{1}$) as  $\alpha_j$ decreases from 1 to 0.

\medskip

\item These two plots (superimposed in the same figure) allow the user to choose a suitable value for $\alpha\in \mathcal{G}$  which is a trade-off between the loss of socio-economic homogeneity and greater geographical cohesion (when viewed through increasing values of $\alpha$) . 
\end{itemize}

\paragraph{Case where the two total (pseudo) inertias $W_0(\mathcal{P}_1)$ and $W_1(\mathcal{P}_1)$ used in $Q_0(\mathcal{P}_K^{\alpha})$ and $Q_1(\mathcal{P}_K^{\alpha})$ are very different.}
Let us consider for instance that the dissimilarity matrix $D_1$ is a ``neighborhood'' dissimilarity matrix, constructed  from the corresponding adjacency matrix $A$: that is $D_1=\mathbf{1}_n-A$ with $\mathbf{1}_{n,ij}=1$ for all $(i,j)$, $a_{ij}$ equal to 1 if observations $i$ and $j$ are neighbors and 0 otherwise, and $a_{ii}=1$ by convention.
With this kind of local dissimilarity matrix $D_1$, the geographical cohesion for few clusters  is often small: indeed, $W_1(\mathcal{P}_1)$ could be very small and thus the criterion $Q_1(\mathcal{P}_K^{\alpha})$ takes values  generally much smaller than those obtained by the $Q_0(\mathcal{P}_K^{\alpha})$. 
Consequently, it is not easy for the user to select easily and properly a suitable value for the mixing parameter $\alpha$ since the two plots are in two very different scales.

One way to remedy this problem is to consider a renormalization of the two plots. 

Rather than reasoning in terms of absolute values of the criterion $Q_0(\mathcal{P}_K^{\alpha})$ (resp. $Q_1(\mathcal{P}_K^{\alpha})$) which is maximal in $\alpha=0$ (resp. $\alpha=1$), we will renormalize $Q_0(\mathcal{P}_K^{\alpha})$ and  $Q_1(\mathcal{P}_K^{\alpha})$ as follows: $Q_0^*(\mathcal{P}_K^{\alpha})=Q_0(\mathcal{P}_K^{\alpha})/Q_0(\mathcal{P}_K^{0})$ and $Q_1^*(\mathcal{P}_K^{\alpha})=Q_1(\mathcal{P}_K^{\alpha})/Q_1(\mathcal{P}_K^{1})$ and we then reason in terms of proportions of these criteria. 
Therefore the corresponding plot $\{(\alpha_j, Q_0^*(\mathcal{P}_K^{\alpha_j})),~j=1,\dots,J\}$ (resp. $\{(\alpha_j, Q_1^*(\mathcal{P}_K^{\alpha_j})),~j=1,\dots,J\}$) 
starts from 100\% and decreases as  $\alpha_j$ increases from 0 to 1 (resp. as $\alpha_j$ decreases from 1 to 0).

\paragraph{The choice of the number $K$ of clusters.}
The proposed procedure to select a suitable value for the mixing parameter $\alpha$ works for a given number $K$ of clusters. Thus, it is first necessary to select $K$.

One way of achieving this is to focus on the dendrogram of the hierarchically-nested set of such partitions $\{\mathcal{P}_n^0=\mathcal{P}_n,\dots,\mathcal{P}_K^0,\dots,\mathcal{P}_1^0=\mathcal{P}_1\}$ 
only based on the dissimilarity matrix $D_0$ (i.e. for $\alpha=0$, that is considering only the socio-economic point of view in our application).
According to the dendrogram, the user can select an appropriate number $K$ of clusters with their favorite rule.

%==================================================================
%==================================================================
\subsection{Description of the  functions of the package {\tt ClustGeo}}
%==================================================================
%==================================================================

The previous Ward-like hierarchical clustering procedure is implemented in the function {\tt hclustgeo} with the following arguments: 
\begin{center}
{\tt hclustgeo(D0, D1 = NULL, alpha = 0, scale = TRUE, wt = NULL)}
\end{center}
where:
\begin{itemize}
\item {\tt D0}  is the dissimilarity matrix $D_0$  between $n$ observations. It must be an object of class {\tt dist}, i.e. an object obtained with the function {\tt dist}.
The function {\tt as.dist} can be used to transform object of class {\tt matrix} to object of class {\tt dist}.
\item {\tt D1} is the dissimilarity matrix $D_1$ between the same $n$ observations. It must be an object of class {\tt dist}. By default {\tt D1=NULL} and the clustering is performed using {\tt D0} only.
\item {\tt alpha} must be a real value between 0 and 1.  The mixing parameter $\alpha$ gives the relative importance of $D_0$ compared to $D_1$. By default, this parameter is equal to 0 and only $D_0$ is used in the clustering process.
\item {\tt scale} must be a logical value. If {\tt TRUE} (by default), the dissimilarity matrices $D_0$ and  $D_1$ are scaled between 0 and 1 (that is divided by their maximum value).
\item {\tt wt} must be a $n$-dimensional vector of the weights of the observations. By default, {\tt wt=NULL} corresponds to the case where all observations are weighted by $1/n$.
\end{itemize}
The function {\tt hclustgeo} returns an object of class {\tt hclust}.

\medskip

\noindent
The  procedure to determine a suitable value for the mixing parameter $\alpha$ is applied through the function {\tt choicealpha} with the following arguments: 
\begin{center}
{\tt choicealpha(D0, D1, range.alpha, K, wt = NULL, scale = TRUE, graph = TRUE)}
\end{center}
where:
\begin{itemize}
\item {\tt D0}  is the dissimilarity matrix $D_0$ of class {\tt dist}, already defined above.
\item {\tt D1} is the dissimilarity matrix $D_1$ of class {\tt dist}, already defined above. 
\item {\tt range.alpha} is the vector of the real values $\alpha_j$ (between 0 and 1) considered by the user in the grid $\mathcal{G}$ of size $J$.
\item {\tt K	} is the number of clusters chosen by the user.
\item {\tt wt} is the vector of the weights of the $n$ observations, already defined above.
\item {\tt scale} is a logical value that allows the user to rescale the dissimilarity matrices $D_0$ and  $D_1$,  already defined above.
\item {\tt graph} is a logical value.	If {\tt graph=TRUE}, the two graphics (proportion and normalized proportion of explained inertia) are drawn.
\end{itemize}
This function returns an object of class {\tt choicealpha} which contains
\begin{itemize}
\item {\tt Q}  is a $J\times 2$ real matrix such that the $j$th row contains $Q_0(\mathcal{P}_K^{\alpha_j})$ and $Q_1(\mathcal{P}_K^{\alpha_j})$.
\item {\tt Qnorm} is a $J\times 2$ real matrix such that the $j$th row contains  $Q_0^*(\mathcal{P}_K^{\alpha_j})$ and $Q_1^*(\mathcal{P}_K^{\alpha_j})$.. 
\item {\tt range.alpha} is the vector of the real values $\alpha_j$ considered in the $\mathcal{G}$.
\end{itemize}
A {\tt plot} method is associated with the class {\tt choicealpha}. 

%%%%%%%%%%%%%%%%%%%%%%%%%%%%%%%%%%%%%%%%%%%%%%%%%%%%%%%%%%
%%%%%%%%%%%%%%%%%%%%%%%%%%%%%%%%%%%%%%%%%%%%%%%%%%%%%%%%%%
%%%%%%%%%%%%%%%%%%%%%%%%%%%%%%%%%%%%%%%%%%%%%%%%%%%%%%%%%%
\section{An illustration of hierarchical clustering with geographical  constraints using the package {\tt ClustGeo}}
\label{sec:4}
%%%%%%%%%%%%%%%%%%%%%%%%%%%%%%%%%%%%%%%%%%%%%%%%%%%%%%%%%%
%%%%%%%%%%%%%%%%%%%%%%%%%%%%%%%%%%%%%%%%%%%%%%%%%%%%%%%%%%
%%%%%%%%%%%%%%%%%%%%%%%%%%%%%%%%%%%%%%%%%%%%%%%%%%%%%%%%%%

This section illustrates the procedure of hierarchical clustering with geographical constraints on a real dataset using the package {\tt ClustGeo}. The complete procedure and methodology for the choice of the mixing parameter $\alpha$ is provided with two types of spatial constraints (with geographical distances and with neighborhood contiguity). We have provided the R code of this case study so that readers can reproduce our methodology and obtain map representations from their own data. 

\subsection{The data} \label{sec:data}

Data were taken from French population censuses conducted by the National Institute of Statistics and Economic Studies (INSEE). The dataset is an extraction of $p = 4$ quantitative socio-economic variables for a subsample of $n = 303$ French municipalities located on the Atlantic coast between Royan and Mimizan:
\begin{itemize}
\item  {\tt employ.rate.city} is the employment rate of the municipality, that is the ratio of the number of individuals who have a job to the population of working age (generally defined, for the purposes of international comparison, as persons of between 15 and 64 years of age). 
\item  {\tt graduate.rate} refers to the level of education of the population, i.e. the highest qualification declared by the individual. It is defined here as the ratio for the whole population having completed a diploma equal to or greater than two years of higher education (DUT, BTS, DEUG, nursing and social training courses, la licence, ma\^itrise, DEA, DESS, doctorate, or Grande Ecole diploma). 
\item  {\tt housing.appart} is the ratio of apartment housing. 
\item  {\tt agri.land} is the part of agricultural area of the municipality.
\end{itemize}

\noindent We consider  here two dissimilarity matrices: 
\begin{itemize} 
\item $D_0$ is the Euclidean distance matrix between the $n$ municipalities performed with the $p=4$ available socio-economic variables, 
\item $D_1$ is a second dissimilarity matrix used to take the geographical proximity between the $n$ municipalities into account. 
\end{itemize} 

{\small
\begin{verbatim}
> library(ClustGeo)
> data(estuary) # list of 3 objects (dat, D.geo, map)  
                # where dat= socio-economic data (n*p data frame), 
                #       D.geo = n*n data frame of geographical distances, 
                #       map = object of class "SpatialPolygonsDataFrame" 
                #             used to draw the map
> head(estuary$dat)
      employ.rate.city graduate.rate housing.appart agri.land
17015            28.08         17.68           5.15  90.04438
17021            30.42         13.13           4.93  58.51182
17030            25.42         16.28           0.00  95.18404
17034            35.08          9.06           0.00  91.01975
17050            28.23         17.13           2.51  61.71171
17052            22.02         12.66           3.22  61.90798
> D0 <- dist(estuary$dat)   # the socio-economic distances
> D1 <- as.dist(estuary$D.geo)   # the geographic distances between the municipalities
\end{verbatim}
}

\subsection{Choice of the number $K$ of clusters}

To choose the suitable number $K$ of clusters, we focus on the Ward dendrogram based on the $p=4$ socio-economic variables, that is using $D_0$ only.

{\small
\begin{verbatim}
> tree <- hclustgeo(D0)
> plot(tree,hang=-1, label=FALSE,  xlab="", sub="",  main="")
> rect.hclust(tree, k=5, border=c(4,  5,  3,  2,  1))
> legend("topright", legend=paste("cluster", 1:5), fill=1:5, bty="n", border="white")
\end{verbatim}
}

\begin{figure}
\centering
\includegraphics[scale=0.4]{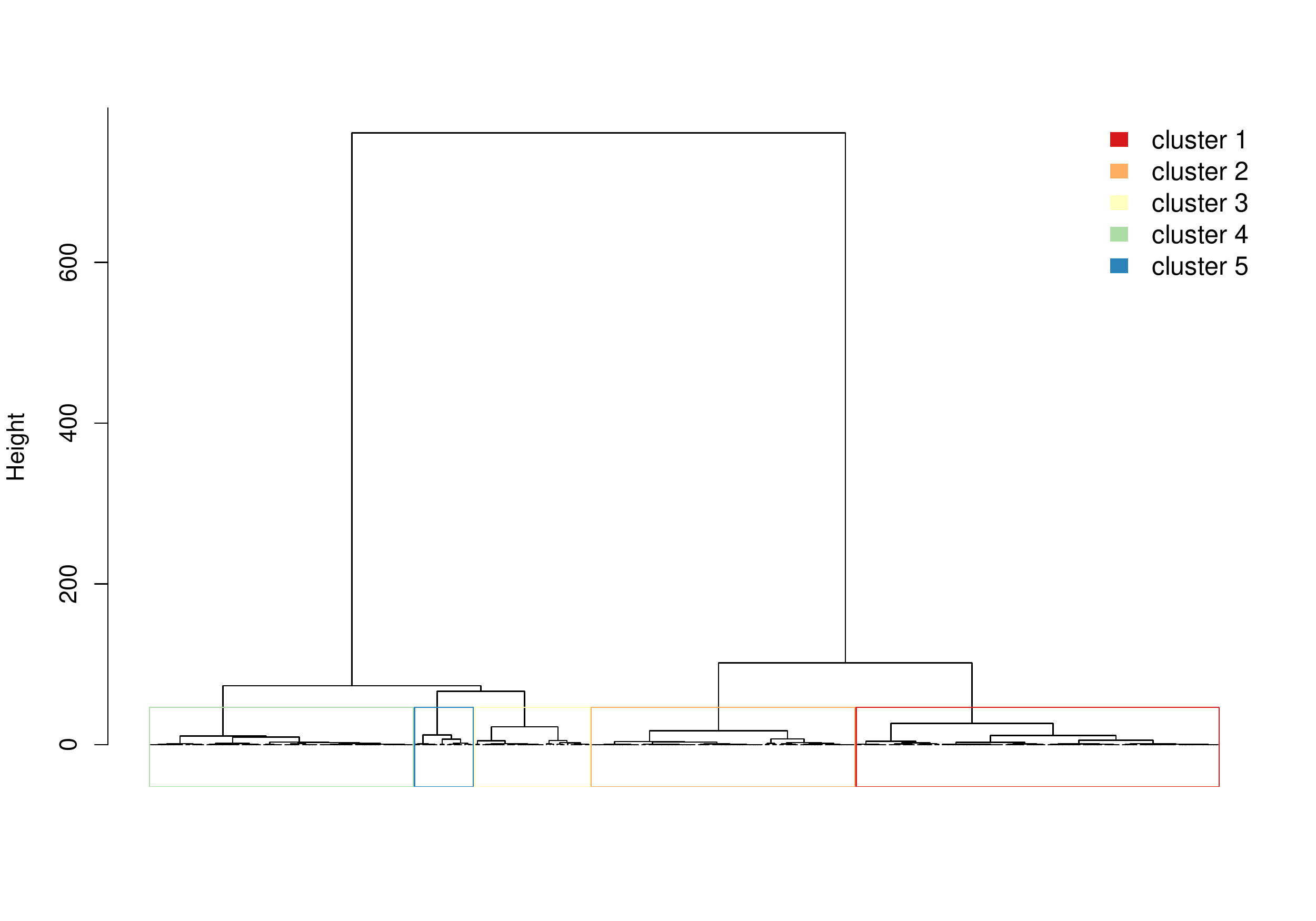}
\vspace{-1cm}
\caption{Dendrogram of the $n=303$ municipalities based on the $p=4$ socio-economic variables (that is using $D_0$ only).}
\label{dendroD0}       
\end{figure}

\noindent
The visual inspection of the dendrogram in Figure~\ref{dendroD0} suggests to retain  $K=5$ clusters.
We can use the map provided in the {\tt estuary} data to visualize the corresponding partition in five clusters, called {\tt P5} hereafter.

{\small
\begin{verbatim}
> P5 <- cutree(tree, 5)  # cut the dendrogram to get the partition in 5 clusters
> sp::plot(estuary$map, border="grey", col=P5) # plot an object of class sp
> legend("topleft", legend=paste("cluster", 1:5), fill=1:5, bty="n", border="white")
\end{verbatim}
}

\begin{figure}
\centering
\includegraphics[scale=0.4]{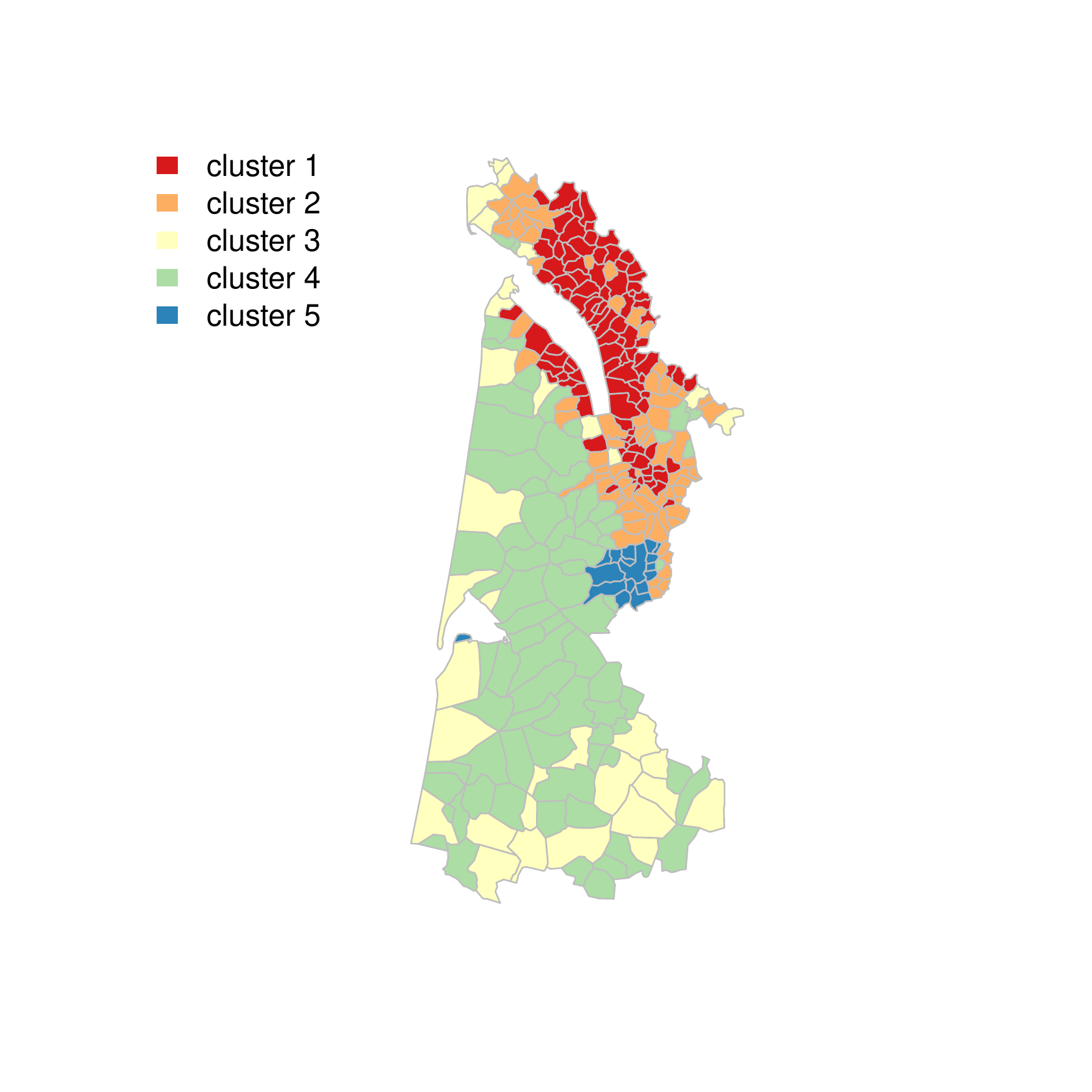}
\vspace{-1cm}
\caption{Map of the partition  {\tt P5} in $K=5$ clusters only based on the socio-economic variables (that is using $D_0$ only).}
\label{map-P5}       
\end{figure}

\noindent
Figure~\ref{map-P5}  shows that municipalities of cluster 5 are geographically compact, corresponding to Bordeaux and the 15 municipalities of its suburban area and Arcachon. On the contrary, municipalities in cluster 3 are scattered over a wider geographical area from North to South of the study area. The composition of each cluster is easily obtained, as shown for cluster 5:

{\small
\begin{verbatim}
# list of the municipalities in cluster 5
> city_label <- as.vector(estuary$map$"NOM_COMM")
> city_label[which(P5 == 5)]
 [1] "ARCACHON"          "BASSENS"           "BEGLES"           
 [4] "BORDEAUX"          "LE BOUSCAT"        "BRUGES"           
 [7] "CARBON-BLANC"      "CENON"             "EYSINES"          
[10] "FLOIRAC"           "GRADIGNAN"         "LE HAILLAN"       
[13] "LORMONT"           "MERIGNAC"          "PESSAC"           
[16] "TALENCE"           "VILLENAVE-D'ORNON"
\end{verbatim}
}

The interpretation of the clusters according to the initial socio-economic variables is interesting. Figure \ref{comparisons}  shows the boxplots of the variables for each cluster of the partition (left column). Cluster~5 corresponds to urban municipalities, Bordeaux and its outskirts plus Arcachon, with a relatively high graduate rate but low employment rate. Agricultural land is scarce and municipalities have a high proportion of apartments. Cluster~2 corresponds to suburban municipalities (north of Royan; north of Bordeaux close to the Gironde estuary) with mean levels of employment and graduates, a low proportion of apartments, more detached properties, and very high proportions of farmland. Cluster~4 corresponds to municipalities located in the Landes forest. Both the graduate rate and the ratio of the number of individuals in employment are high (greater than the mean value of the study area). The number of apartments is quite low and the agricultural areas are higher to the mean value of the zone. Cluster~1 corresponds to municipalities on the banks of the Gironde estuary. The proportion of farmland is higher than in the other clusters. On the contrary, the number of apartments is the lowest. However this cluster also has both the lowest employment rate and the lowest graduate rate. Cluster~3 is geographically sparse. It has the highest employment rate of the study area, a graduate rate similar to that of cluster~2, and a collective housing rate equivalent to that of cluster 4. The agricultural area is low. 

\subsection{Obtaining a partition taking into account the geographical constraints}

To obtain more geographically compact clusters, we can now introduce the matrix $D_1$ of geographical distances into {\tt hclustgeo}. 
This requires a mixing parameter to be selected $\alpha$ to improve the geographical cohesion of the 5 clusters without adversely affecting socio-economic cohesion.

\paragraph{Choice of the mixing parameter $\alpha$.}
The mixing parameter $\alpha \in [0,1]$ sets the importance of $D_0$ and $D_1$ in the clustering process. When $\alpha=0$ the geographical dissimilarities are not taken into account and when $\alpha=1$ it is the socio-economic distances which are not taken into account and the clusters are obtained with the geographical distances only. 

\noindent
The idea is to perform separate calculations for socio-economic homogeneity and the geographic cohesion of the partitions obtained for a range of different values of $\alpha$ and a given number of clusters $K$.

\noindent
To achieve this, we can plot the quality criterion $Q_0$ and $Q_1$ of the partitions $P_K^\alpha$ obtained with different values of $\alpha \in [0,1]$ and choose the value of $\alpha$ which is a trade-off between the lost of socio-economic homogeneity and the gain of geographic cohesion. We use the function {\tt choicealpha} for this purpose. 

{\small
\begin{verbatim}
> cr <- choicealpha(D0, D1, range.alpha=seq(0, 1, 0.1), K=5, graph=TRUE)
> cr$Q # proportion of explained pseudo-inertia
                 Q0        Q1
alpha=0   0.8134914 0.4033353
alpha=0.1 0.8123718 0.3586957
alpha=0.2 0.7558058 0.7206956
alpha=0.3 0.7603870 0.6802037
alpha=0.4 0.7062677 0.7860465
alpha=0.5 0.6588582 0.8431391
alpha=0.6 0.6726921 0.8377236
alpha=0.7 0.6729165 0.8371600
alpha=0.8 0.6100119 0.8514754
alpha=0.9 0.5938617 0.8572188
alpha=1   0.5016793 0.8726302
> cr$Qnorm # normalized proportion of explained pseudo-inertias
             Q0norm    Q1norm
alpha=0   1.0000000 0.4622065
alpha=0.1 0.9986237 0.4110512
alpha=0.2 0.9290889 0.8258889
alpha=0.3 0.9347203 0.7794868
alpha=0.4 0.8681932 0.9007785
alpha=0.5 0.8099142 0.9662043
alpha=0.6 0.8269197 0.9599984
alpha=0.7 0.8271956 0.9593526
alpha=0.8 0.7498689 0.9757574
alpha=0.9 0.7300160 0.9823391
alpha=1   0.6166990 1.0000000
\end{verbatim}
}

\begin{figure}[h]
\centering
\includegraphics[scale=0.4]{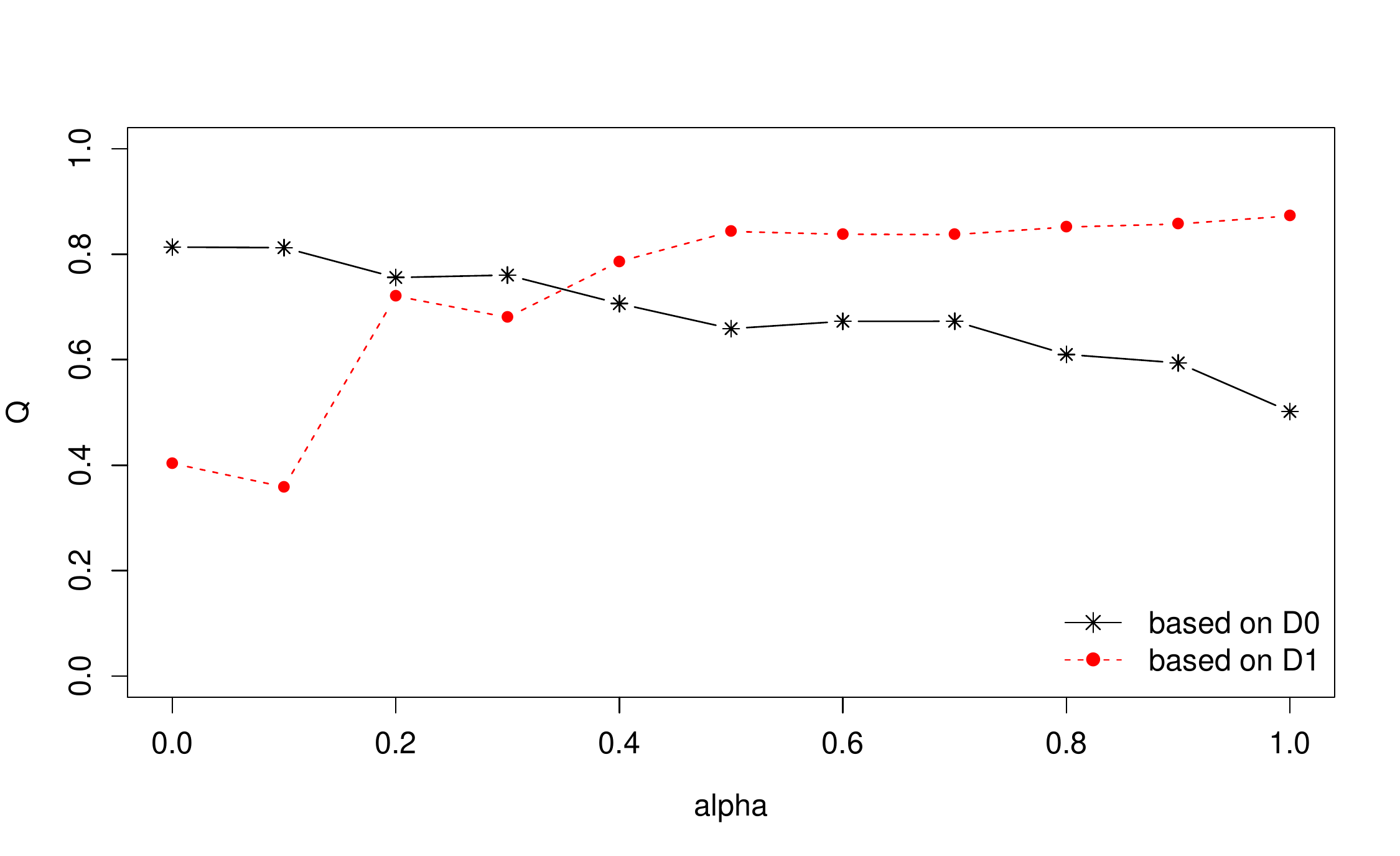}
\includegraphics[scale=0.4]{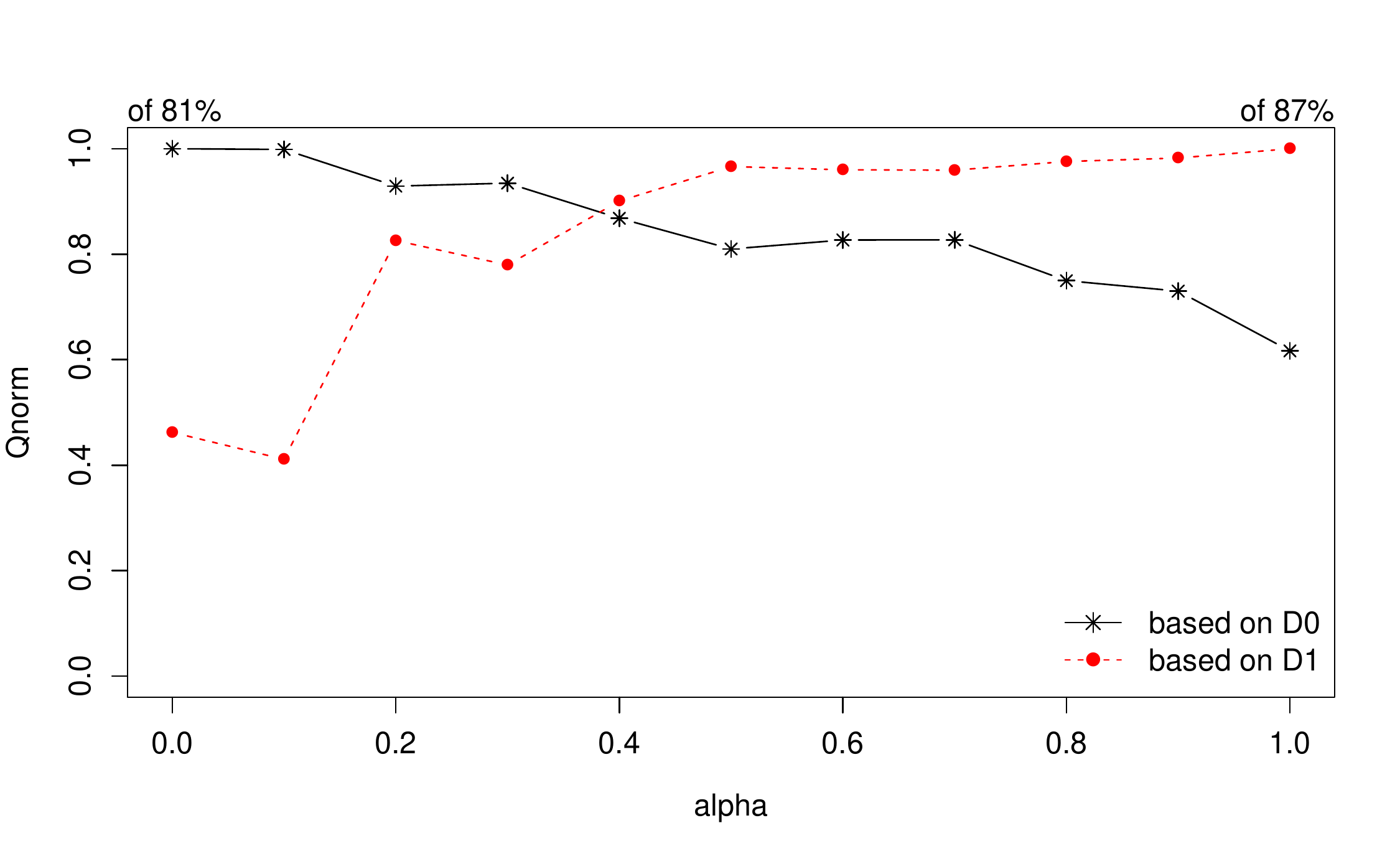}
\caption{Choice of $\alpha$ for a partition in $K=5$ clusters when $D1$ is the geographical distances between municipalities. Top: proportion of explained pseudo-inertias $Q_0(\mathcal{P}_K^{\alpha})$ versus $\alpha$ (in black solid line) and $Q_1(\mathcal{P}_K^{\alpha})$ versus $\alpha$ (in dashed line). Bottom: normalized proportion of explained pseudo-inertias  $Q_0^*(\mathcal{P}_K^{\alpha})$ versus $\alpha$ (in black solid line) and $Q_1^*(\mathcal{P}_K^{\alpha})$ versus $\alpha$ (in dashed line).}
\label{plotQ-1}       
\end{figure}

\noindent
Figure~\ref{plotQ-1} gives the plot of  the proportion of explained pseudo-inertia calculated with $D_0$ (the socio-economic distances) which is equal to 0.81 when $\alpha=0$ and decreases when $\alpha$ increases (black solid line).  On the contrary, the proportion of explained pseudo-inertia calculated with $D_1$ (the geographical distances) is equal to 0.87 when  $\alpha=1$ and decreases when $\alpha$ decreases (dashed line).

\noindent
Here, the plot would appear to suggest choosing $\alpha=0.2$ which corresponds to a loss of only 7\% of socio-economic homogeneity, and a 17\% increase in geographical homogeneity. 

\paragraph{Final partition obtained with $\alpha=0.2$.}
We perform {\tt hclustgeo} with $D_0$ and $D_1$ and $\alpha=0.2$ and cut the tree to get the new partition in five clusters, called {\tt P5bis} hereafter.

{\small
\begin{verbatim}
> tree <- hclustgeo(D0, D1, alpha=0.2)
> P5bis <- cutree(tree, 5)
> sp::plot(estuary$map, border="grey", col=P5bis)
> legend("topleft", legend=paste("cluster", 1:5), fill=1:5, bty="n", border="white")
\end{verbatim}
}

\begin{figure}[h]
\centering
\includegraphics[scale=0.4]{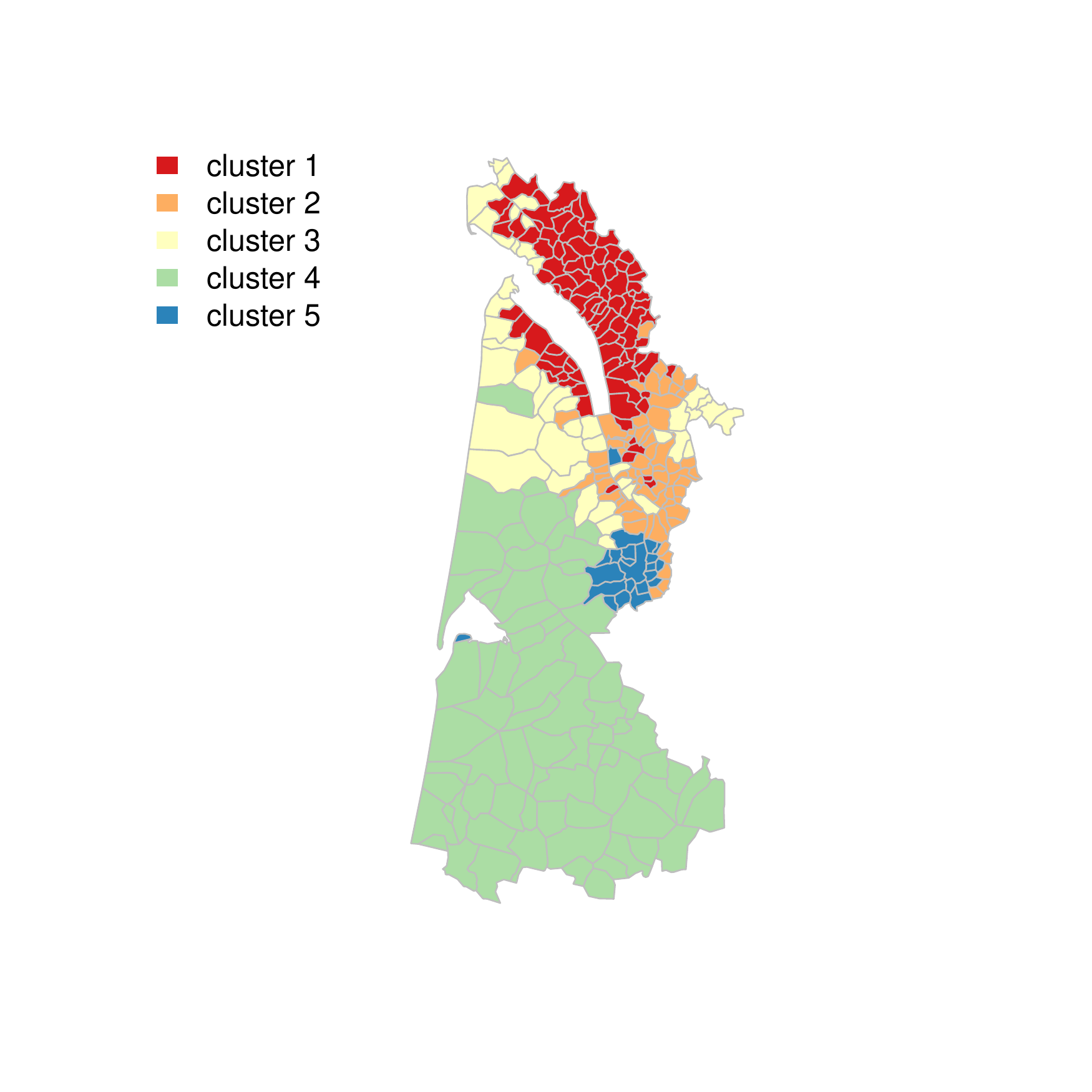}
\vspace{-1cm}
\caption{Map of the partition  {\tt P5bis} in $K=5$ clusters based on the socio-economic distances $D_0$ and  the geographical distances between the municipalities $D_1$ with $\alpha=0.2$.}
\label{map-P5bis}       
\end{figure}

\noindent
The increased geographical cohesion of this partition  {\tt P5bis} can  be seen in Figure~\ref{map-P5bis}. Figure~\ref{comparisons} shows the boxplots of the variables for each cluster of the partition   {\tt P5bis}  (middle column). Cluster~5 of  {\tt P5bis} is  identical to cluster~5 of {\tt P5} with the Blaye municipality added in. Cluster~1 keeps the same interpretation as in {\tt P5} but has gained spatial homogeneity. It is now clearly located on the banks of the Gironde estuary, especially on the north bank. The same applies for cluster~2 especially for municipalities between Bordeaux and the estuary. Both clusters 3 and 4 have changed significantly. Cluster~3 is now a spatially compact zone, located predominantly in the M\'edoc. 

It would appear that these two clusters have been separated based on proportion of farmland, because the municipalities in cluster~3 have above-average proportions of this type of land, while cluster~4 has the lowest proportion of farmland of the whole partition. Cluster~4 is also different because of the increase in clarity both from a spatial and socio-economic point of view. In addition, it contains the southern half of the study area. The ranges of all variables are also lower in the corresponding boxplots.

\subsection{Obtaining a partition taking into account the neighborhood constraints}
Let us construct a different type of matrix $D_1$ to take neighbouring municipalities into account when clustering the 303 municipalities.   

Two regions with contiguous boundaries, that is sharing one or more boundary point, are considered as neighbors.  Let us first build the adjacency matrix $A$.

{\small
\begin{verbatim}
> list.nb <- spdep::poly2nb(estuary$map,
                            row.names=rownames(estuary$dat)) #list of neighbors
\end{verbatim}
}

\noindent
It is possible to obtain the list of the neighbors of a specific city. For instance, the neighbors of Bordeaux (which is the 117th city in the R data table) is given by the script:

{\small
\begin{verbatim}
> city_label[list.nb[[117]]] # list of the neighbors of BORDEAUX 
 [1] "BASSENS"     "BEGLES"      "BLANQUEFORT" "LE BOUSCAT"  "BRUGES"     
 [6] "CENON"       "EYSINES"     "FLOIRAC"     "LORMONT"     "MERIGNAC"   
[11] "PESSAC"      "TALENCE"    
\end{verbatim}
}

\noindent
The dissimilarity matrix $D_1$ is constructed based on the adjacency matrix $A$ with $D_1=\mathbf{1}_n-A$. 

{\small
\begin{verbatim}
> A <- spdep::nb2mat(list.nb, style="B")   # build the adjacency matrix
> diag(A) <- 1
> colnames(A) <- rownames(A) <- city_label
> D1 <- 1-A
> D1[1:2, 1:5]
         ARCES ARVERT BALANZAC BARZAN BOIS
ARCES        0      1        1      0    1
ARVERT       1      0        1      1    1
> D1 <- as.dist(D1)
\end{verbatim}
}

\paragraph{Choice of the mixing parameter $\alpha$.}
The same procedure for the choice of $\alpha$ is then used with this neighborhood dissimilarity matrix $D_1$.

{\small
\begin{verbatim}
> cr <- choicealpha(D0, D1, range.alpha=seq(0, 1, 0.1), K=5, graph=TRUE) 
> cr$Q # proportion of explained pseudo-inertia
> cr$Qnorm # normalized proportion of explained pseudo-inertia
\end{verbatim}
}

%\vspace{-0.4cm}

\begin{figure}[h!]
\centering
\includegraphics[scale=0.4]{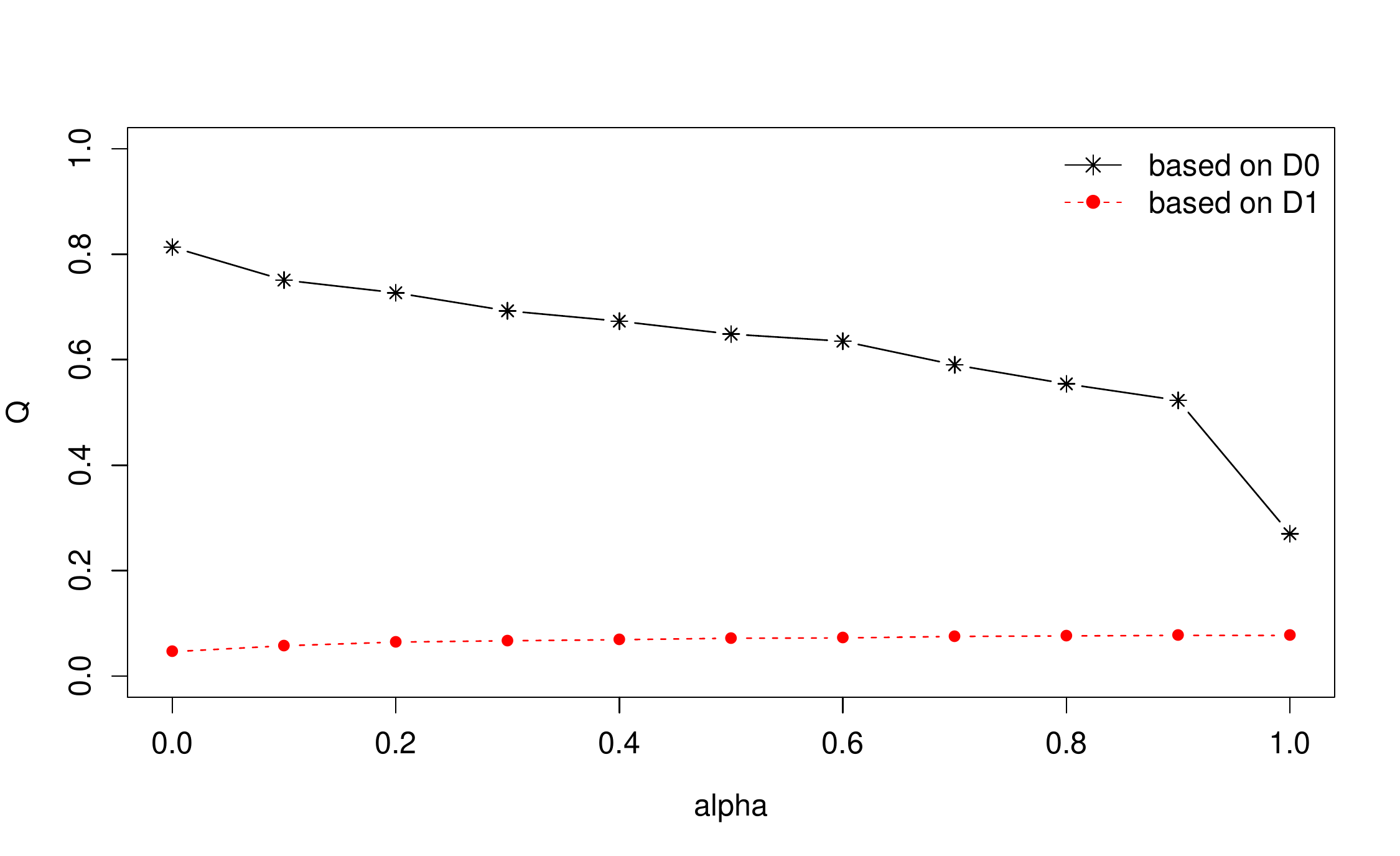}
\includegraphics[scale=0.4]{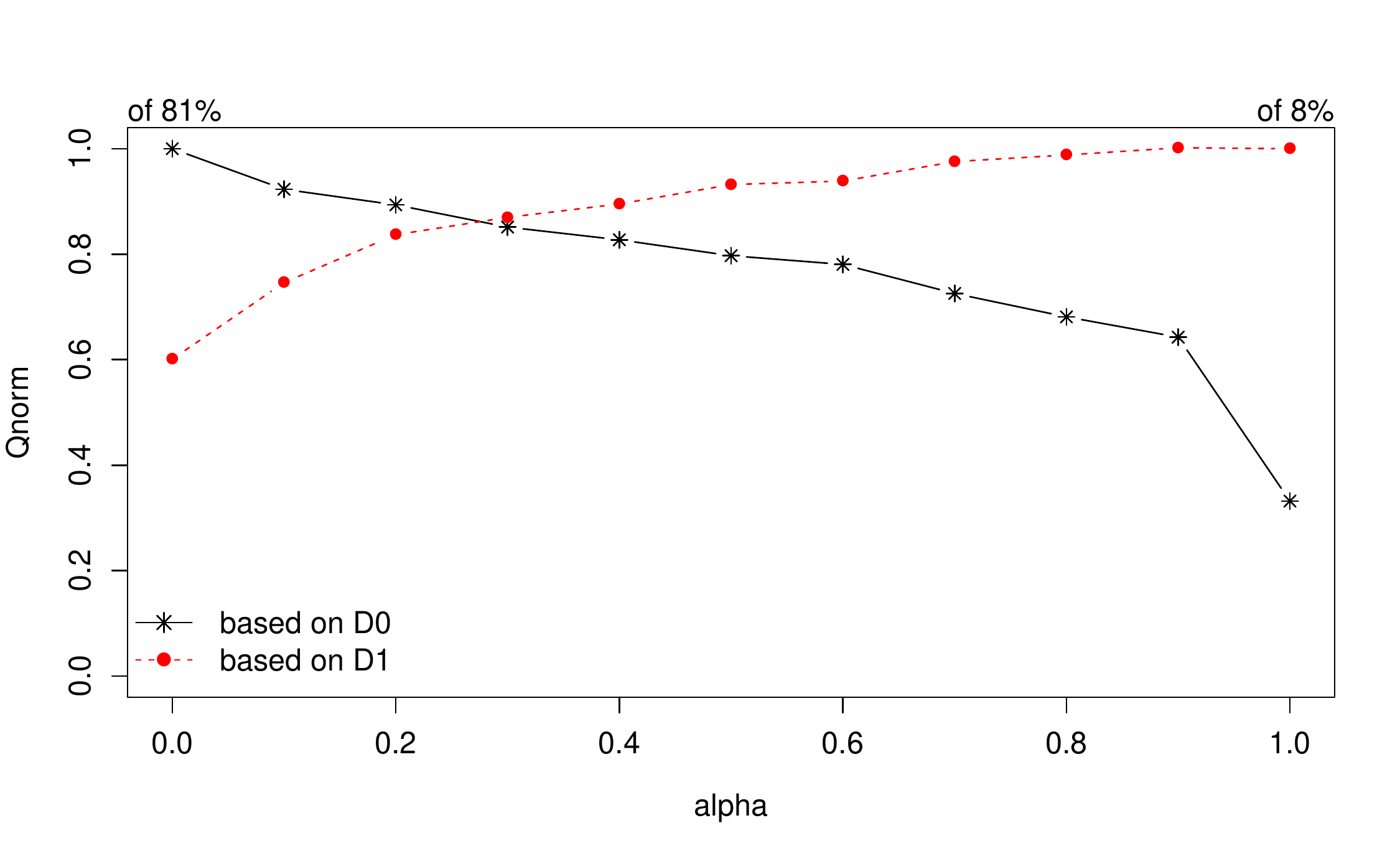}
\caption{Choice of $\alpha$ for a partition in $K=5$ clusters when $D1$ is the neighborhood dissimilarity matrix between municipalities. Top: proportion of explained pseudo-inertias $Q_0(\mathcal{P}_K^{\alpha})$ versus $\alpha$ (in black solid line) and $Q_1(\mathcal{P}_K^{\alpha})$ versus $\alpha$ (in dashed line). Bottom: normalized proportion of explained pseudo-inertias  $Q_0^*(\mathcal{P}_K^{\alpha})$ versus $\alpha$ (in black solid line) and $Q_1^*(\mathcal{P}_K^{\alpha})$ versus $\alpha$ (in dashed line).}
\label{plotQ-2}       
\end{figure}

\noindent
With these kinds of local dissimilarities in $D_1$, the neighborhood within-cluster cohesion  is always very small. $Q_1(\mathcal{P}_K^{\alpha})$ takes small values: see the dashed line of $Q_1(\mathcal{P}_K^{\alpha})$ versus $\alpha$ at the top of Figure~\ref{plotQ-2}. To overcome this problem, the user can plot the normalized proportion of explained inertias (that is $Q_0^*(\mathcal{P}_K^{\alpha})$ and $Q_1^*(\mathcal{P}_K^{\alpha})$) instead of the proportion of explained inertias (that is $Q_0(\mathcal{P}_K^{\alpha})$ and $Q_1(\mathcal{P}_K^{\alpha})$). At the bottom of Figure~\ref{plotQ-2}, the plot of the normalized proportion of explained inertias suggests  to retain $\alpha=0.2$ or $0.3$. The value $\alpha=0.2$ slightly favors the socio-economic homogeneity versus the geographical homogeneity. According to the priority given in this application to the socio-economic aspects, the final partition is obtained with $\alpha=0.2$.

\begin{figure}[h]
\centering
\includegraphics[scale=0.4]{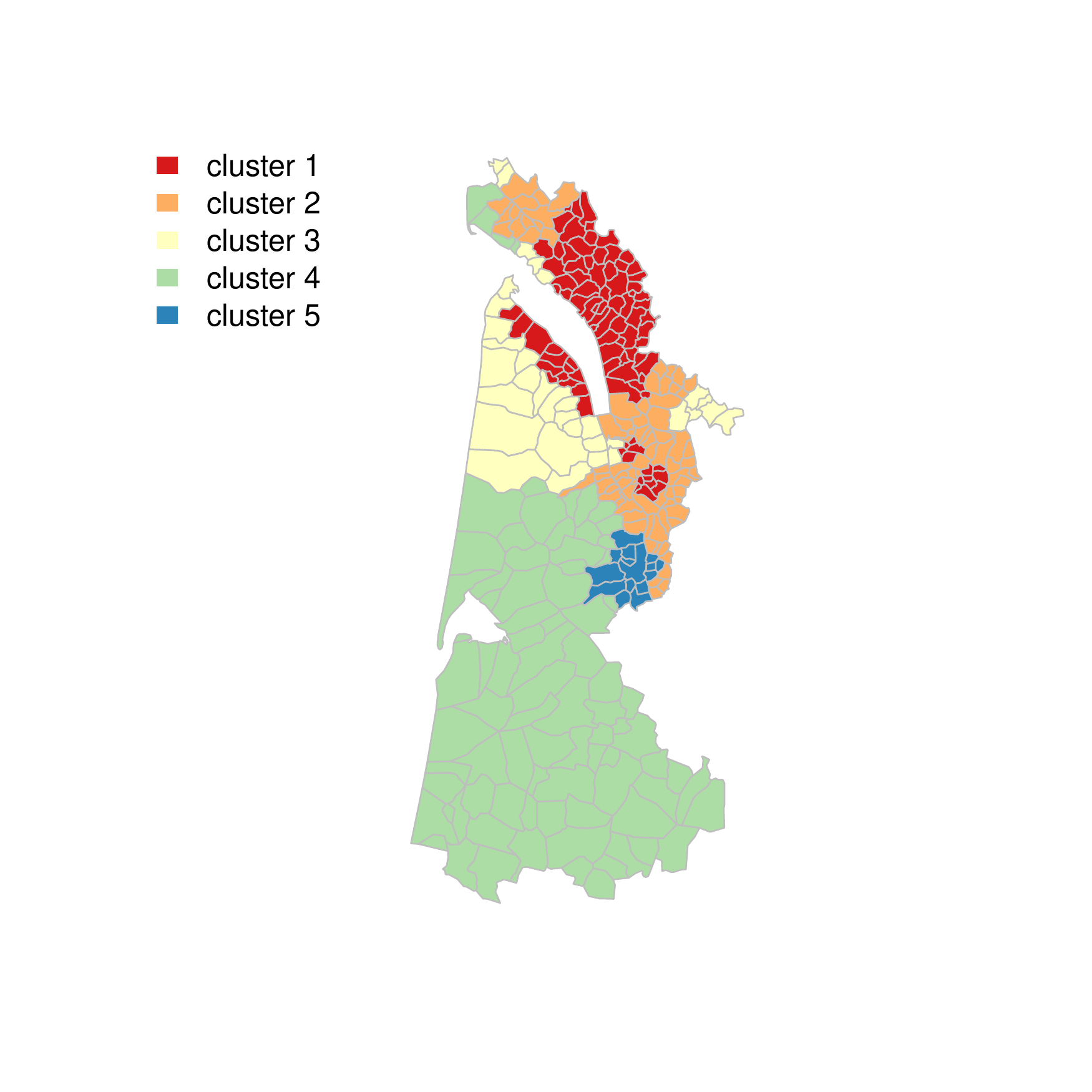}
\vspace{-1cm}
\caption{Map of the partition {\tt P5ter} in $K=5$ clusters based on the socio-economic distances $D_0$ and  the ``neighborhood'' distances of the municipalities $D_1$ with $\alpha=0.2$.}
\label{map-P5ter}       
\end{figure}

\paragraph{Final partition obtained with $\alpha=0.2$.}
It remains only to determine this final partition for $K=5$ clusters and $\alpha = 0.2$,  called {\tt P5ter} hereafter. The  corresponding map is given in Figure~\ref{map-P5ter}.

{\small
\begin{verbatim}
> tree <- hclustgeo(D0, D1, alpha=0.2)
> P5ter <- cutree(tree, 5)
> sp::plot(estuary$map, border="grey", col=P5ter)
> legend("topleft", legend=paste("cluster", 1:5), fill=1:5, bty="n", border="white")
\end{verbatim}
}

%\vspace{-0.4cm}

\begin{figure}
\centering
\includegraphics[scale=0.35]{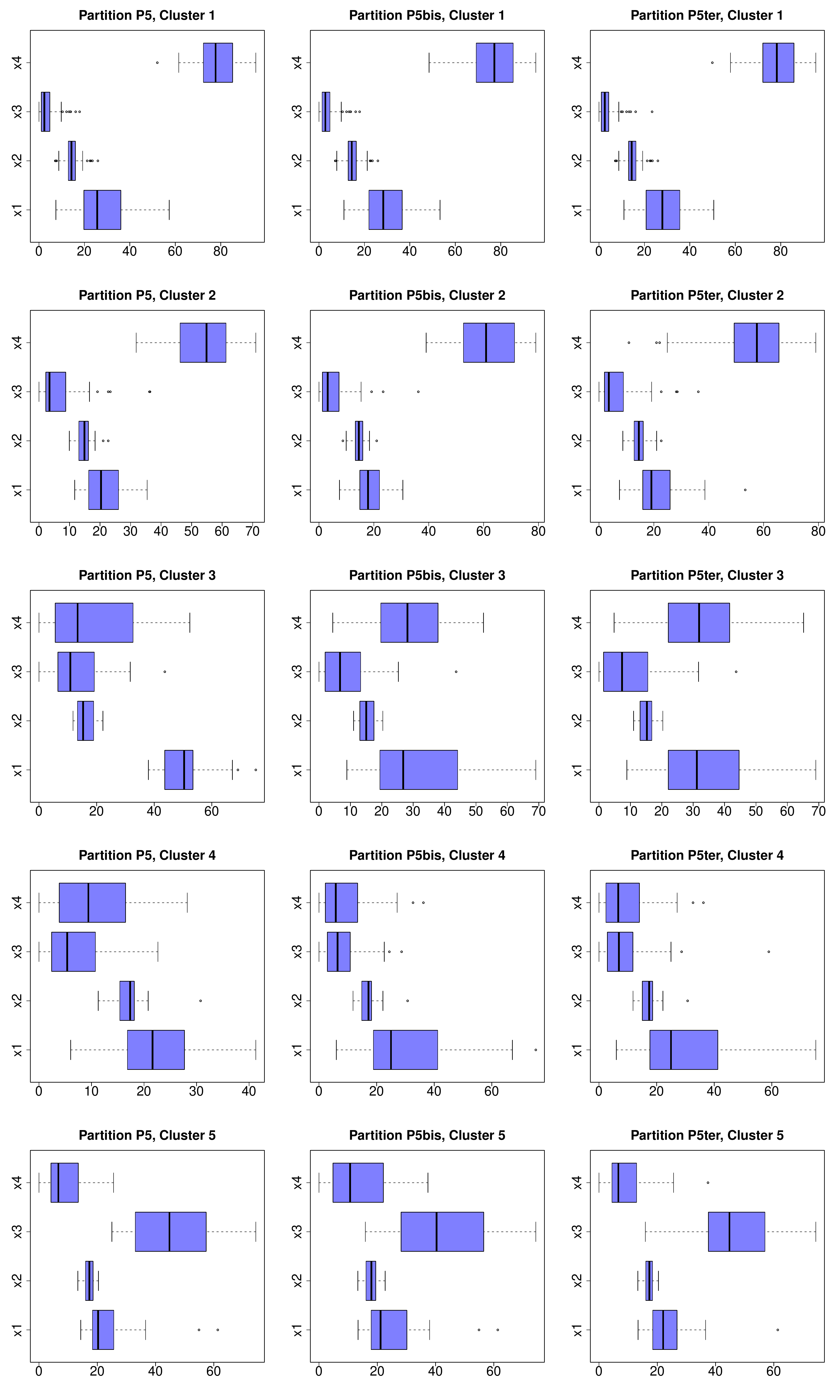}

\caption{Comparison of the final partitions {\tt P5}, {\tt P5bis} and {\tt P5ter} in terms of variables  x1=employ.rate.city, x2=graduate.rate,  x3=housing.appart  and x4=agri.land.}
\label{comparisons}       
\end{figure}

\noindent
Figure \ref{map-P5ter} shows that clusters of {\tt P5ter} are spatially more compact than that of {\tt P5bis}. This is not surprising since this approach builds  dissimilarities from the adjacency matrix which gives more importance to neighborhoods. However since our approach is based on soft contiguity constraints,  municipalities that are not neighbors are allowed to be in the same clusters. This is the case for instance for cluster~4  where some municipalities are located in the north of the estuary whereas  most are located in the southern area (corresponding to forest areas). The quality of the partition {\tt P5ter} is  slightly worse than that  of partition  {\tt P5ter} according to criterion $Q_0$  (72.69\% versus  75.58\%).  However the boxplots corresponding to partition {\tt P5ter}  given in Figure \ref{comparisons} (right column)  are very similar to those of partition {\tt P5bis}. These two partitions have thus very close interpretations.

%%%%%%%%%%%%%%%%%%%%%%%%%%%%%%%%%%%%%%%%%%%%%%%%%%%%%%%%%%
%%%%%%%%%%%%%%%%%%%%%%%%%%%%%%%%%%%%%%%%%%%%%%%%%%%%%%%%%%
%%%%%%%%%%%%%%%%%%%%%%%%%%%%%%%%%%%%%%%%%%%%%%%%%%%%%%%%%%
\section{Concluding remarks}
\label{sec:5}
%%%%%%%%%%%%%%%%%%%%%%%%%%%%%%%%%%%%%%%%%%%%%%%%%%%%%%%%%%
%%%%%%%%%%%%%%%%%%%%%%%%%%%%%%%%%%%%%%%%%%%%%%%%%%%%%%%%%%
%%%%%%%%%%%%%%%%%%%%%%%%%%%%%%%%%%%%%%%%%%%%%%%%%%%%%%%%%%

In this paper, a Ward-like hierarchical clustering algorithm including soft spacial constraints has been introduced and illustrated on a real dataset. 
The corresponding approach has been implemented in the R package {\tt ClustGeo} available on the CRAN. When the observations correspond to geographical units (such as a city or a region), it is then possible to represent the clustering obtained on a map regarding the considered spatial constraints. This Ward-like hierarchical clustering method can also be used in many other contexts where the  observations do not correspond to geographical units. In that case, the dissimilarity matrix $D_1$ associated with the ``constraint space'' does not correspond to spatial constraints in its current form. 

For instance, the user may have at his/her disposal a first set of data of $p_0$ variables (e.g. socio-economic items) measured on $n$ individuals on which he/she has made a clustering from the associated dissimilarity (or distance) matrix. This user also has a second data set of $p_1$ new variables (e.g. environmental items) measured on these same $n$ individuals, on which a dissimilarity matrix $D_1$ can be calculated. Using the {\tt ClusGeo} approach, it is possible to take this new information into account to refine the initial clustering without basically disrupting it.

\section*{References}

{\small
\begin{itemize}

\item[] \hspace{-0.8cm}  Ambroise C, Dang M, Govaert G (1997) Clustering of spatial data by the EM algorithm.  In: Soares A, G\`omez-Hernandez J, Froidevaux R (eds) geoENV~I~-Geostatistics for Environmental Applications. Springer, pp. 493-504

\item[] \hspace{-0.8cm}  Ambroise C, Govaert G  (1998) Convergence of an EM-type algorithm for spatial clustering. Pattern Recognition Letters 19(10): 919-927 

\item[] \hspace{-0.8cm}   B\'ecue-Bertaut  M, Alvarez-Esteban R, S\`anchez-Espigares JA (2017) Xplortext: Statistical Analysis of Textual Data R package. \url{https://cran.r-project.org/package=Xplortext}. R package version 1.0

\item[] \hspace{-0.8cm}   B\'ecue-Bertaut  M, Kostov B, Morin A , Naro G (2014)  Rhetorical strategy in forensic speeches: multidimensional statistics-based methodology. Journal of Classication 31(1): 85-106 

\item[] \hspace{-0.8cm}  Bourgault G, Marcotte D, Legendre P (1992) The Multivariate (co) Variogram as a Spatial Weighting Function in Classification Methods. Mathematical Geology 24(5): 463-478 

\item[] \hspace{-0.8cm}  Chavent M, Kuentz-Simonet V, Labenne A, Saracco J (2017) ClustGeo: Hierarchical Clustering with Spatial Constraints. \url{https://cran.r-
project.org/package=ClustGeo}. R package version 2.0

\item[] \hspace{-0.8cm}   Dehman A, Ambroise C, Neuvial P (2015) Performance of a blockwise approach in variable selection using linkage disequilibrium information. BMC Bioinformatics 16:148. 

\item[] \hspace{-0.8cm}  Duque JC, Dev B, Betancourt A, Franco JL (2011) ClusterPy: Library of spatially constrained clustering algorithms,  RiSE-group (Research in Spatial Economics). EAFIT University. \url{http://www.rise-group.org/risem/clusterpy/}. Version 0.9.9. 

\item[] \hspace{-0.8cm} Ferligoj A,  Batagelj V (1982) Clustering with relational constraint. Psychometrika 47(4):413-426 

\item[] \hspace{-0.8cm} Gordon AD (1996) A survey of constrained classication. Computational Statistics \& Data Analysis 21:17-29 

\item[] \hspace{-0.8cm} Lance GN, Williams WT (1967) A General Theory of Classicatory Sorting Strategies 1. Hierarchical Systems. The Computer Journal 9:373-380

\item[] \hspace{-0.8cm} Legendre P  (2014) const.clust: Space-and Time-Constrained Clustering Package. \url{http://adn.biol.umontreal.ca/ numericalecology/Rcode/}

\item[] \hspace{-0.8cm}  Legendre P, Legendre L  (2012) Numerical Ecology, vol. 24. Elsevier 

\item[] \hspace{-0.8cm} Miele V, Picard F, Dray S (2014) Spatially constrained clustering of ecological networks. Methods in Ecology and Evolution 5(8):771-779

\item[] \hspace{-0.8cm} Murtagh F (1985a) Multidimensional clustering algorithms. Compstat Lectures, Vienna: Physika Verlag 

\item[] \hspace{-0.8cm} Murtagh F (1985b) A Survey of Algorithms for Contiguity-constrained Clustering and Related Problems. The Computer Journal 28:82-88 

\item[] \hspace{-0.8cm} Oliver M, Webster R (1989) A Geostatistical Basis for Spatial Weighting in Multivariate Classication. Mathematical Geology 21(1):15-35 

\item[] \hspace{-0.8cm} Strauss T, von Maltitz MJ (2017) Generalising Ward's Method for Use with Manhattan Distances. PloS ONE 12(1). \url{https://doi.org/10.1371/journal.pone.0168288}

\item[] \hspace{-0.8cm} Vignes M, Forbes F (2009) Gene Clustering via Integrated Markov Models Combining Individual and Pairwise Features. IEEE/ACM Transactions on Computational Biology and Bioinformatics (TCBB) 6(2):260-270 

\item[] \hspace{-0.8cm} Ward Jr JH  (1963) Hierarchical Grouping to Optimize an Objective Function. Journal of the American Statistical Association 58(301):236-244

\end{itemize}

}
%=======================
%=======================
%=======================
\end{document}